\documentclass[aps,twocolumn,amsmath,amssymb,floatfixng,superscriptaddress,longbibliography,numerical]{revtex4-1}
\pdfoutput=1
\usepackage[dvips]{graphics}
\usepackage{bm}
\usepackage{epsfig}
\usepackage{enumerate}
\usepackage{subfigure}
\usepackage{color}
\usepackage{graphicx}
\usepackage{hyperref}
\hypersetup{
    pdfnewwindow=true,      
    colorlinks=true,       
    linkcolor=magenta,          
    citecolor=blue,        
    filecolor=blue,      
    urlcolor=blue        
}

\newcommand{\bea}{\begin{eqnarray}}
\newcommand{\eea}{\end{eqnarray}}
\newcommand{\beq}{\begin{equation}}  
\newcommand{\eeq}{\end{equation}}

\begin{document} 

\title{Photoinduced spin-Hall resonance in a $k^3$-Rashba spin-orbit coupled two-dimensional hole system} 

\author{Ankita Bhattacharya}
\email{ankita.bhattacharya@tu-dresden.de}
\affiliation{Institute of Theoretical Physics, Technische Universit\"at Dresden, 01062 Dresden, Germany}

\author{SK Firoz Islam}
\email{firoz.seikh@aalto.fi}
\affiliation{Department of Applied Physics, Aalto University, P.~O.~Box 15100, FI-00076 AALTO, Finland}

\begin{abstract}
We study the band structure modulation and spin-Hall effect of a two-dimensional heavy-hole system with $k^3$-Rashba spin-orbit coupling (RSOC), irradiated by linearly polarized light. We find that the band structure becomes anisotropic under the illumination by the light. Most remarkably, a pair of additional spin-degeneracy points (apart from $\Gamma$ point) emerge in the energy dispersion, the locations of which are solely determined by the strength of the amplitude of the incident light. If this degeneracy occurs around the Fermi level, the spin-Hall conductivity (SHC) exhibits a resonance. Away from the degeneracy points, the light rotates the average spin polarization. The possible effects of $k^3$-Dresselhaus spin-orbit coupling are also discussed.  
\end{abstract}
\maketitle
{\em Introduction.}
Recent years have witnessed growing interest in the interaction of light with electronic systems, especially after the discovery of light-induced topological phase 
transitions\cite{PhysRevB.79.081406,PhysRevB.85.125425,PhysRevB.89.235416,PhysRevLett.110.026603,PhysRevB.89.121401,cayssol2013floquet} (Floquet topological insulator), 
followed by several experiments \cite{lindner2011floquet,peng2016experimental,zhang2014anomalous,wang2013observation}. Such interaction can also be used to manipulate the 
spin- and valley-selective transport in Dirac materials\cite{PhysRevB.85.205428,PhysRevLett.116.016802,PhysRevB.84.195408,PhysRevB.98.075422} and transition-metal 
dichalcogenides\cite{PhysRevB.90.125438,tahir2014tunable}. 
There have been extensive works on photo tunable Weyl nodes\cite{PhysRevLett.117.087402,PhysRevB.94.235137,PhysRevB.97.155152,PhysRevB.94.041409,PhysRevB.96.041205}, 0-$\pi$ 
transition in Josephson junction\cite{PhysRevB.94.165436,PhysRevB.95.201115}, and Floquet engineering of Majorana 
modes\cite{PhysRevLett.111.136402,PhysRevLett.111.047002,PhysRevLett.126.086801,PhysRevB.99.094303}. 
The interfacial chiral modes have also been predicted in threefold topological semimetals\cite{PhysRevB.100.165302}, by controlling the phase of the light. 
More recently, higher-order topological insulators\cite{PhysRevB.100.085138,PhysRevB.101.235403,PhysRevResearch.2.013124} have also been predicted in different irradiated  materials.
 
On the other hand, two-dimensional systems with strong spin-orbit interactions have become promising test beds for future spintronics development.
The RSOC\cite{Winkler,RevModPhys.76.323} 
in 2D fermionic systems arises due to the lack of structural inversion symmetry in the quantum well of semiconductor heterojunctions, which removes spin 
degeneracy in the absence of a magnetic field. Several experiments have confirmed that the strength of RSOC can be significantly enhanced by applying a gate voltage across the quantum
well\cite{PhysRevB.55.R1958,PhysRevLett.78.1335,PhysRevB.57.11911}. There are two types of RSOC, namely linear and cubic in momentum($\mathrm{k}$). Typical materials which exhibit $k$-linear RSOC are 
indium based compounds such as InAs, GaInAs/GaAlAs structure, and II-VI semiconductor compounds. On the other hand, $k^3$-RSOC arises due to the
structural bulk inversion asymmetry for heavy holes in the quantum well of III-V semiconductor heterojunction. The {\it Spin Hall Effect} (SHE)\cite{PhysRevLett.83.1834,PhysRevLett126603,
RevModPhys.87.1213,murakami2003} is one of the exciting phenomena in a 2D electron/hole gas with RSOC, in which  electrons or holes with opposite
spin are accumulated at opposite transverse edges normal to the applied in-plane electric field. The spin-Hall conductivity (SHC) was predicted to exhibit a universal constant value in 2D electron gas ($2$DEG) with $k$-RSOC\cite{PhysRevLett126603}. However, further study revealed that the disorder induced vortex correction strongly suppresses the SHC in such systems\cite{PhysRevB.70.041303,dimitrova2004}. The SHC in a $k^3$-RSOC 2D heavy-hole gas($2$DHG) was also investigated and found to be robust to the vortex correction\cite{PhysRevB.69.241202}, 
for which this system has an edge over $2$DEG with RSOC. Moreover, the strong spin-orbit coupling and weak hyperfine interaction in a heavy-hole gas allows low-power electrical manipulation of 
spin\cite{PhysRevLett.95.076805,PhysRevLett.118.016801,li2015pauli}. Another study on the SHE in a 2D hole system has revealed that the SHC  strongly depends on the strength  of RSOC\cite{PhysRevB.71.085308}, 
which is in contrast to the corresponding electron system. The magnetic-field-dependent transport signatures (Shubnikov-de Hass oscillation and quantum-Hall conductivity)  
\cite{PhysRevB.41.8278,PhysRevB.67.085313,Islam2011,Mawrie_2014,Mawrie_2017,PhysRevB.98.155442} have been considered in both systems, revealing the roles of RSOC.  Recently, a series of theoretical 
works on spin-related phenomena in $2$DHG in presence of weak magnetic field have been studied by the Culcer group\cite{PhysRevLett.121.087701,
PhysRevB.101.121302,PhysRevLett.121.077701,cullen.2020}. Moreover, the search for the materials or mechanisms to achieve giant or large SHC\cite{Zhueaav8025,PhysRevB.101.064430,PhysRevB.101.094435} continues to be an active topic in the field of spintronics. It is noteworthy that 
the SHE is one of the key signatures of spin-Hall edge modes in the quantum-spin-Hall liquid discovered in 2D Dirac materials with spin-orbit 
coupling\cite{PhysRevLett.95.226801,PhysRevLett.109.055502}. However, due to the restricted number of spin-Hall edge modes and unavoidable bulk effects, the enhancement of SHC is 
challenging there. Moreover, from an experiment perspective, semiconductor compounds are still much more reliable platform than 2D Dirac materials. 

The study of interplay of the light with $k$-RSOC in 2DEG has been initiated by Ojanen et.al.\cite{PhysRevB.85.161202}, predicting chiral edge states and out-of-plane magnetization. Very recently, the combined effects of light and real magnetic field have also been studied in this system\cite{PhysRevB.102.165414}. However, to the best of our 
knowledge, the effects of light on 2DHG in the presence of $k^3$-RSOC  has not been addressed yet. 

\begin{figure}
\includegraphics[width=8cm,height=5.5cm]{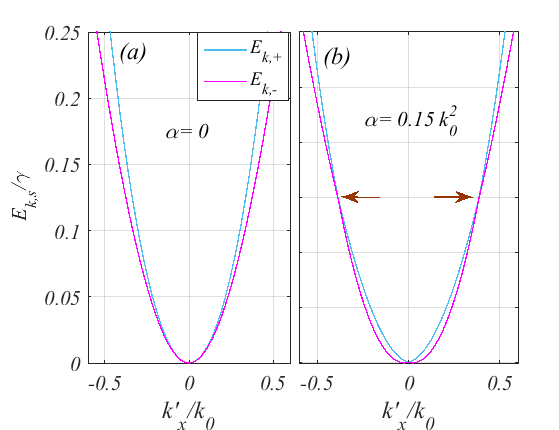}
\caption{Energy dispersions are shown for  (a) $\alpha=0$ and (b) $\alpha=0.15 k_0^2$. Here, momentum axis is rotated by $\pi/4$ ( $k'_x=k_x/\sqrt{2})$ to capture the degeneracy points which are shown by two arrows. The energy axis is normalized by $\gamma=\hbar^2k_0^2/2m^{\ast}$, where $k_0$ is a typical wave vector, corresponds to standard hole density. The strength of RSOC is taken to be $\beta k_0^3=0.4$.}
\label{band}
\end{figure}

In this work, we consider the interaction of linearly polarized light with 2DHG in the presence of $k^3$-RSOC. The irradiated band structure is found to be anisotropic. Instead of out-of-plane magnetization, linearly polarized light rotates the average spin polarization in the plane of 2DEG. Most remarkably, we reveal that 
there exist two spin-degeneracy points in the band dispersion, which give rise to a resonance in the SHC. We also discuss the possible effects of the $k^3$-Dresselhaus spin-orbit coupling (DSOC) on SHC.\\
{\em Model Hamiltonian and energy spectrum.}
 We start with a $2$DHG in the presence of $k^3$-RSOC. The hole dynamics at the top of the valence band in III-V semiconductor quantum wells is generally described by the $4\times 4$ Luttinger Hamiltonian\cite{PhysRev.102.1030}. The large splitting between the heavy-hole(HH) states ($\mid 3/2,\pm 3/2\rangle$) and the light-hole(LH) states ($\mid 3/2,\pm 1/2\rangle$) and due to the higher density of states of HH states, contribution in the transport properties comes predominantly from the HH states close to the Fermi energy. Thus the model system can be described by a $2\times2$ Hamiltonian by projecting onto the HH states.
 The single-particle Hamiltonian of this system is given by\cite{PhysRevB.71.085308,PhysRevB.62.4245} $H=\hbar^2k^2/2m^{\ast}+H_R$ where the RSOC is given by 
 $
  H_R=(\beta/2 i)(k_{-}^3\sigma_{+}-k_{+}^3\sigma_{-})
 $
with $k_{\pm}=k_x\pm ik_y$ and $\sigma_{\pm}=\sigma_{x}\pm i\sigma_{y}$. Here, ${\mathbf{ k}}=(k_x,k_y)$ are the $2$D momentum operators whereas ${\bf \sigma}=(\sigma_x,\sigma_y)$ are the two of the three Pauli spin matrices which act on the total angular momentum states with spin projection $\pm 3/2$ , respectively. The strength of the RSOC and the effective mass of the hole are denoted by $\beta$ and $m^{\ast}$, respectively. The spin-dependent energy  spectrum can be written as $E_{k,s}=\hbar^2k^2/2m^{\ast}+s\beta\mid k\mid^3$, where $s=\pm$ stands for two different spin branches. The dispersion is shown in Fig.\ref{band}(a). It is to be noted that this model Hamiltonian is valid when the wave number $ k\le\hbar^2/2m^{\ast}\beta$. 

{\em Effects of irradiation.}
Let us now consider that the system is subjected to an external time-dependent periodic perturbation in the form of irradiation(light), propagating along the $z$-direction. The light field is described by a vector potential $\mathbf{A}(t)=A_0[\sin(\Omega t), \sin(\Omega t+\phi_0)]$, where $A_0$ is the field amplitude, $\Omega$ is the frequency of the 
irradiation, and $\phi_0$ is the phase. Here, we consider that the light wavelength is large compare to the lattice constant, for which we can safely ignore the spatial dependence of the light field. The light-induced vector field can be included into the Hamiltonian as $\mathbf{k}\rightarrow \mathbf{k}+e\mathbf{A}(t)/\hbar$, where $e<0$ is the electron charge. 
To solve the periodically perturbed Hamiltonian, we can use the Floquet theory\cite{RevModPhys.89.011004}, which states that such a perturbed Hamiltonian exhibits a complete set of orthonormal solutions of the form  $\psi(t)=\phi(t)e^{-iE t/\hbar}$ with $\phi(t)=\phi(t+\mathcal{T})$ being the corresponding Floquet states and $\mathcal{T}$  is the period of the field. 
Here, $E$ denotes the Floquet quasi-energy. The time-dependent Schr{\"o}dinger  equation corresponding to the Floquet states yields the Floquet eigenvalue equation as $H_{F}\phi(t)=E\phi(t)$ with $H_{F} = H(t)-i\hbar \partial_t$. The Floquet states can be further expressed as $\phi(t)=\sum_{n}\phi{_n}(t)e^{in\Omega t}$, where $n$ is the Fourier component or Floquet side-band index. By diagonalizing the Floquet Hamiltonian in the basis of Floquet side bands `$n$', one can obtain the Floquet quasi-energy spectrum. On the other hand, an  effective Hamiltonian can be obtained in the high-frequency limit following the Floquet-Magnus expansion in power of $1/(\hbar\Omega)$ as $H_{F}\simeq H+H_F^{(0)}+H_F^{(1)}+\ldots$, where
 $H_F^{(0)}=-\alpha\beta\cos\phi_0(\sigma_x k_x+\sigma_y k_y)$
with $\alpha=3(eA_0/\hbar)^2$ and $H_{F}^{(1)}=[H_{-},H_{+}]/\hbar\Omega.$ Here,  
\begin{equation}
 H_{m}=\frac{1}{\mathcal{T}}\int_0^{\mathcal{T}}V(t)e^{-im\Omega t}dt
\end{equation}
with $m=\pm$. Here, $V(t) ={\bf h(t)\cdot\sigma}$, which for linearly polarized light ($\phi_0=0$) simplifies to
\begin{eqnarray}
 h_x(t)&=&\beta(eA_0/\hbar)[3(k_y^2-k_x^2-2k_xk_y)\sin(\Omega t)\nonumber\\&-&2(eA_0/\hbar)^2\sin^3(\Omega t)+3(eA_0/\hbar)k_x\cos(2\Omega t)]
\end{eqnarray}
and
\begin{eqnarray}
 h_y(t)&=&\beta (eA_0/\hbar)[3(k_x^2-k_y^2-2k_xk_y)\sin(\Omega t)\nonumber\\&-&2(eA_0/\hbar)^2\sin^3(\Omega t)+3(eA_0/\hbar)k_y\cos(2\Omega t)].
\end{eqnarray}
 The high frequency limit in our case is described as $(\gamma, \beta k_0^3,e{A}_0\hbar k_{0}/m^{\ast}, 3e{A}_0\beta k_{0}^2/\hbar,e^2{A}_0^2\beta k_{0}/\hbar^2)<<\hbar{\Omega}$.
The first-order correction term $H_{F}^{(1)}$ vanishes for linearly polarized light. Other higher-order terms are too small to be considered here. Hence, the irradiated Hamiltonian can be simplified to an effective Hamiltonian as 
$
 H_{eff}\simeq H + H_F^{(0)},
$
which can be diagonalized to obtain the Floquet energy spectrum as 
\begin{equation}
 E_{k,s}=\frac{\hbar^2 k^2}{2m^{\ast}}+s\beta k\sqrt{[k^2-\alpha\sin(2\theta)]^2+\alpha^2\cos^2(2\theta)}
\end{equation}
where $\tan\theta=k_y/k_x$. Note that the Floquet energy spectrum does not depend on the frequency of irradiation. The energy dispersion is now anisotropic for $\alpha\neq 0$. This is one of the interesting effects of irradiation. The corresponding eigenstates are given by
$
 \mid k,s\rangle=\left[\begin{array}{cc}
                1&
    ise^{-i\chi}\end{array}
\right]^{T}e^{i(k_xx+k_yy)}/\sqrt{2}
$
where $
 \tan\chi=[\alpha\cos\theta-k^2\sin(3\theta)]/[\alpha\sin\theta+k^2\cos(3\theta)].
$
Another interesting observation here is that the energy splitting between the two spin branches vanishes at the $(k_r,\theta_{r})=(\sqrt{\alpha},\theta_{r})$, when $\theta_r=(\pi/4,5\pi/4)$, i.e., apart from $k=0$, a pair of additional spin-degeneracy points appears. Note that the locations of these degeneracy points do not depend on the strength of RSOC explicitly. The degeneracy points are shown by arrows in the Flqouet energy spectrum, plotted in Fig.~\ref{band}(b). If this degeneracy occurs at the Fermi level then SHC exhibits a resonance, known as spin-Hall resonance(SHR) which was first discovered in a 2D electronic system with RSOC in the presence of a constant magnetic 
field\cite{PhysRevLett.92.256603}. The competition between Zeeman splitting and Rashba splitting causes such a degeneracy between two nearest Landau levels with opposite spin. The SHR phenomena was studied immediately in the $2$DHG with $k^3$-RSOC\cite{doi:10.1063/1.2345024}. However, note that in both works, the magnetic field plays the key role. In our work, we find that such a spin degeneracy can be achieved even without Landau levels, by applying linearly polarized light. Before going to discuss the SHC, we shall briefly discuss the effects of light on the average spin polarization.\\ 
 
{\em Spin polarization.}
To examine the effects of irradiation on average spin polarization in projected spin space, we can evaluate different components of the average spin polarization as 
\begin{equation}
 \langle k,s\mid \hat{S}_x\mid k, s\rangle=s\frac{k^2\sin(3\theta)-\alpha\cos\theta}{\sqrt{k^4+\alpha^2-2\alpha k^2\sin(2\theta)}}
\end{equation}
and
\begin{equation}
 \langle k,s\mid \hat{S}_y\mid k, s\rangle=s\frac{k^2\cos(3\theta)+\alpha\sin \theta}{\sqrt{k^4+\alpha^2-2\alpha k^2\sin(2\theta)}}  
\end{equation}
where $\hat{S}_{x,y}=(3/2)\hbar\hat{\sigma}_{x,y}$.
\begin{figure}
\centering
\begin{minipage}[t]{.5\textwidth}
\hspace{-.7cm}{\includegraphics[height=4cm,width=5cm]{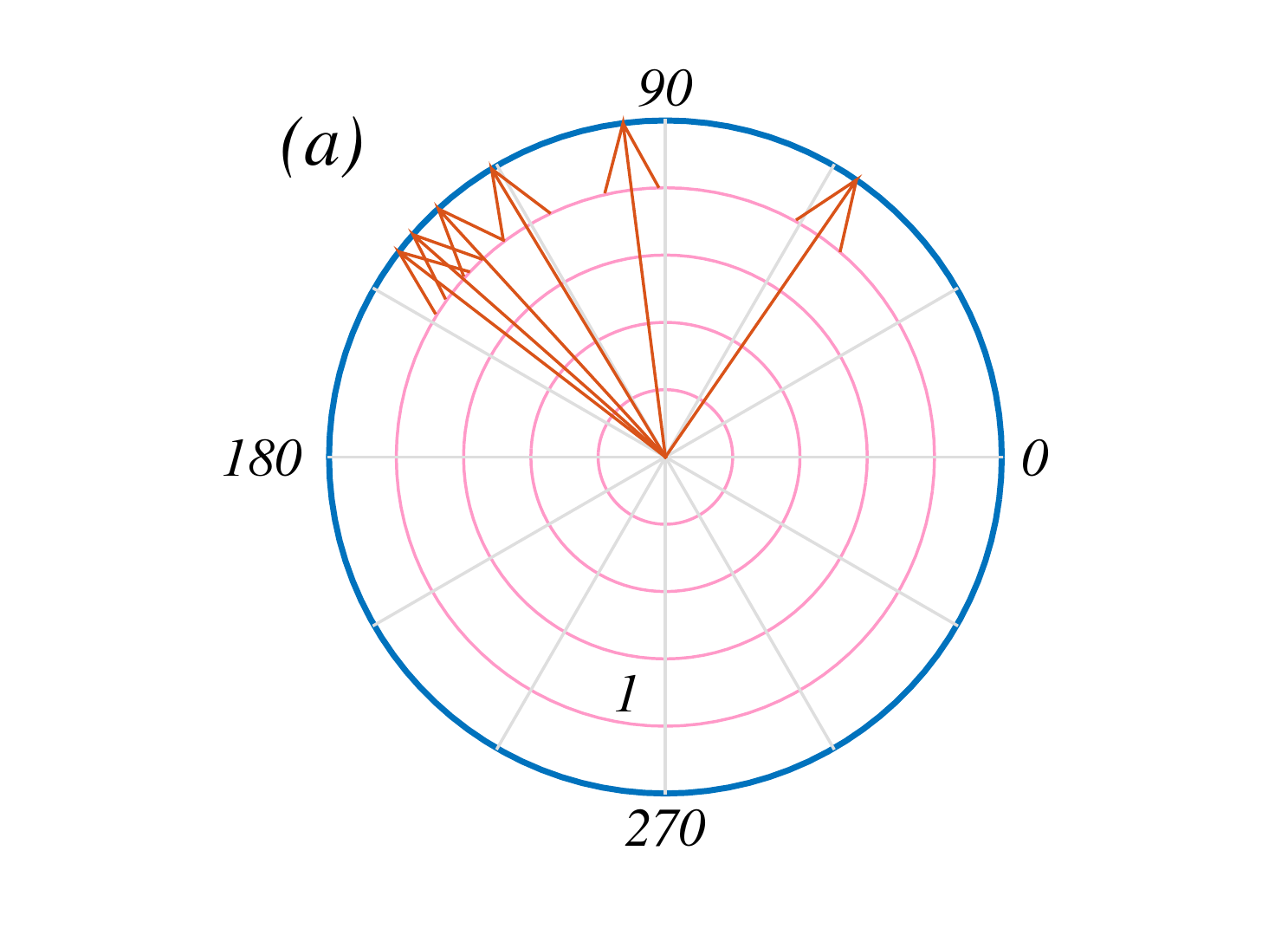}}
\hspace{-.9cm}{\includegraphics[height=4cm,width=5cm]{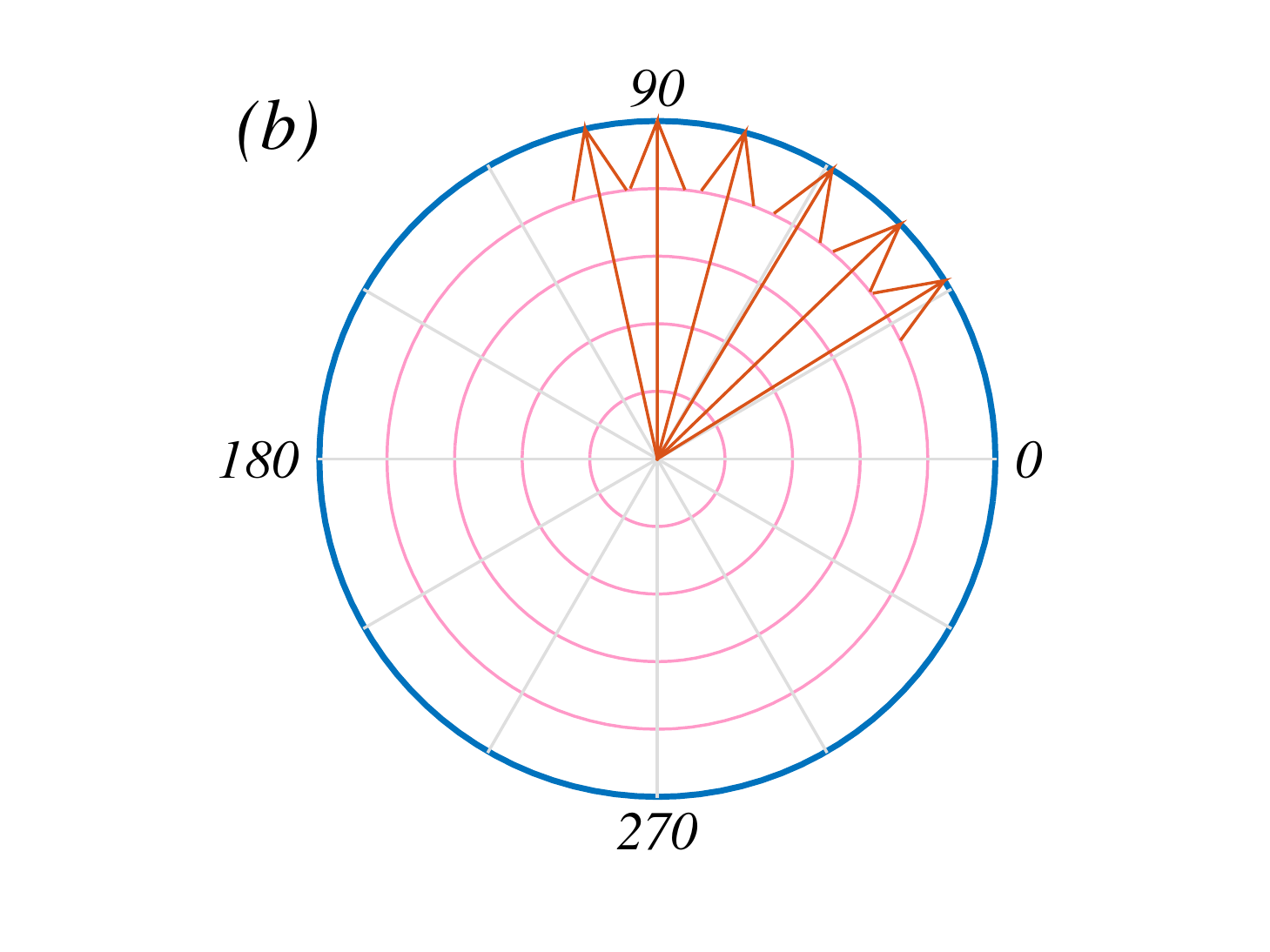}} 
\end{minipage} 
\caption{Spin rotations in $\langle S_x\rangle-\langle S_y\rangle$ plane are shown for (a) $k/k_0=0.3$, $\theta=\pi/8$  and (b) $k/k_0=0.5$, $\theta=\pi/8$ with increasing $\alpha/k_0^2$ from $0.05$ to $0.3$ in steps of $0.05$. Spin rotates in anticlockwise direction with increasing $\alpha$. Both components of average spin are in units of $(3/2)\hbar$.}
\label{CP}
\end{figure}
It can be realised from the above expressions that away from the spin degeneracy point, the spin polarization is significantly affected by the strength of irradiation, $\alpha$. The effects of light on the average spin polarization is presented by a couple of compass plots in Fig.~(\ref{CP}), showing anticlockwise rotation of the average spin with the increase of $\alpha$. We consider two sets of $(k,\theta)$, showing the sensitivity of spin rotation on momentum. It is to be noted that in the $k$-RSOC electron system\cite{PhysRevB.85.161202}, the effects of light on spin orientation or band structure can be observed only for circularly polarized light, whereas in our case the light is linearly polarized.\\

\begin{figure}[h!]
\centering
\begin{minipage}[t]{.5\textwidth}
 \hspace{-.45cm}{ \includegraphics[width=.52\textwidth,height=4.5cm]{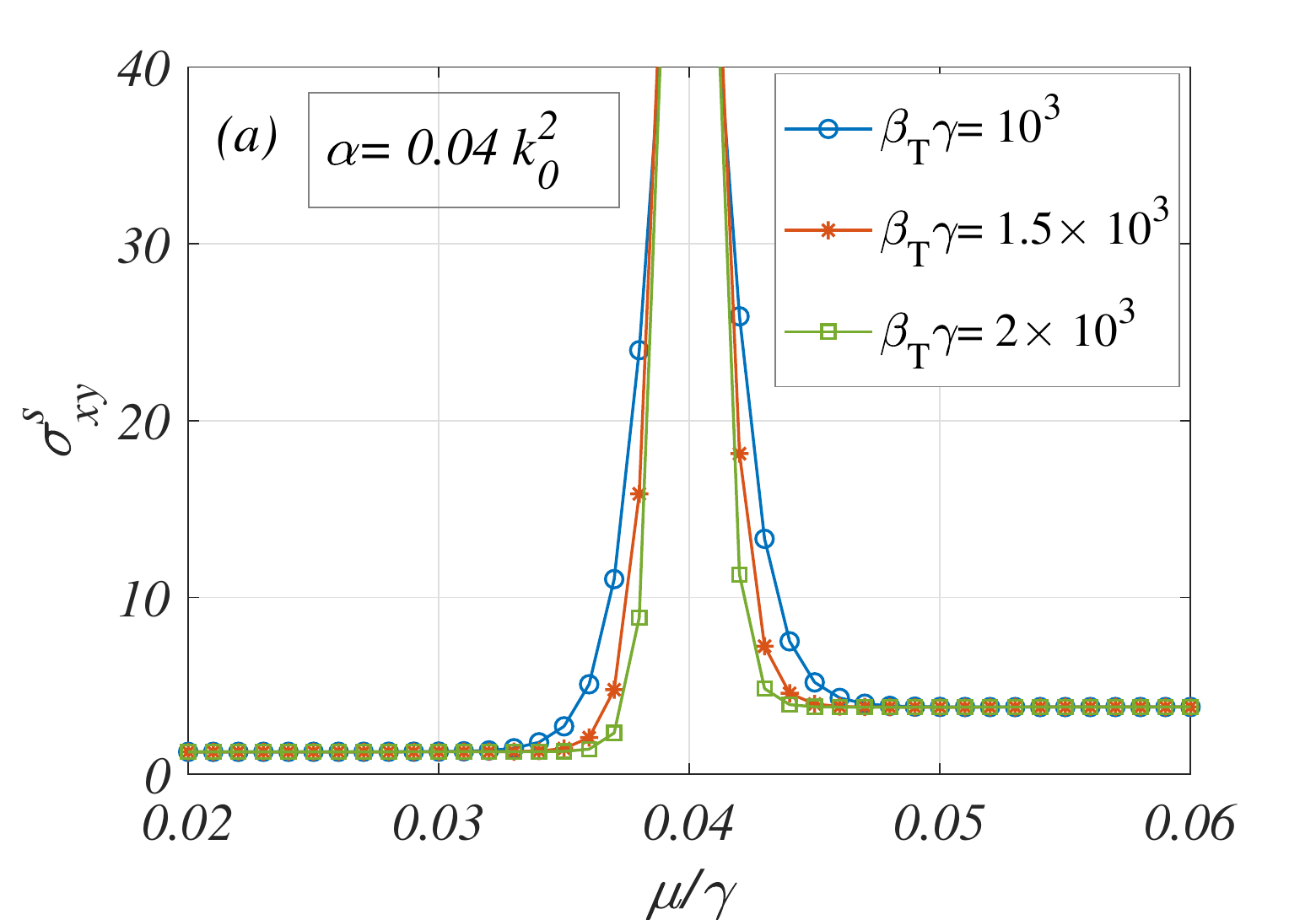}}
  \hspace{-0.42cm}{ \includegraphics[width=.48\textwidth,height=4.5cm]{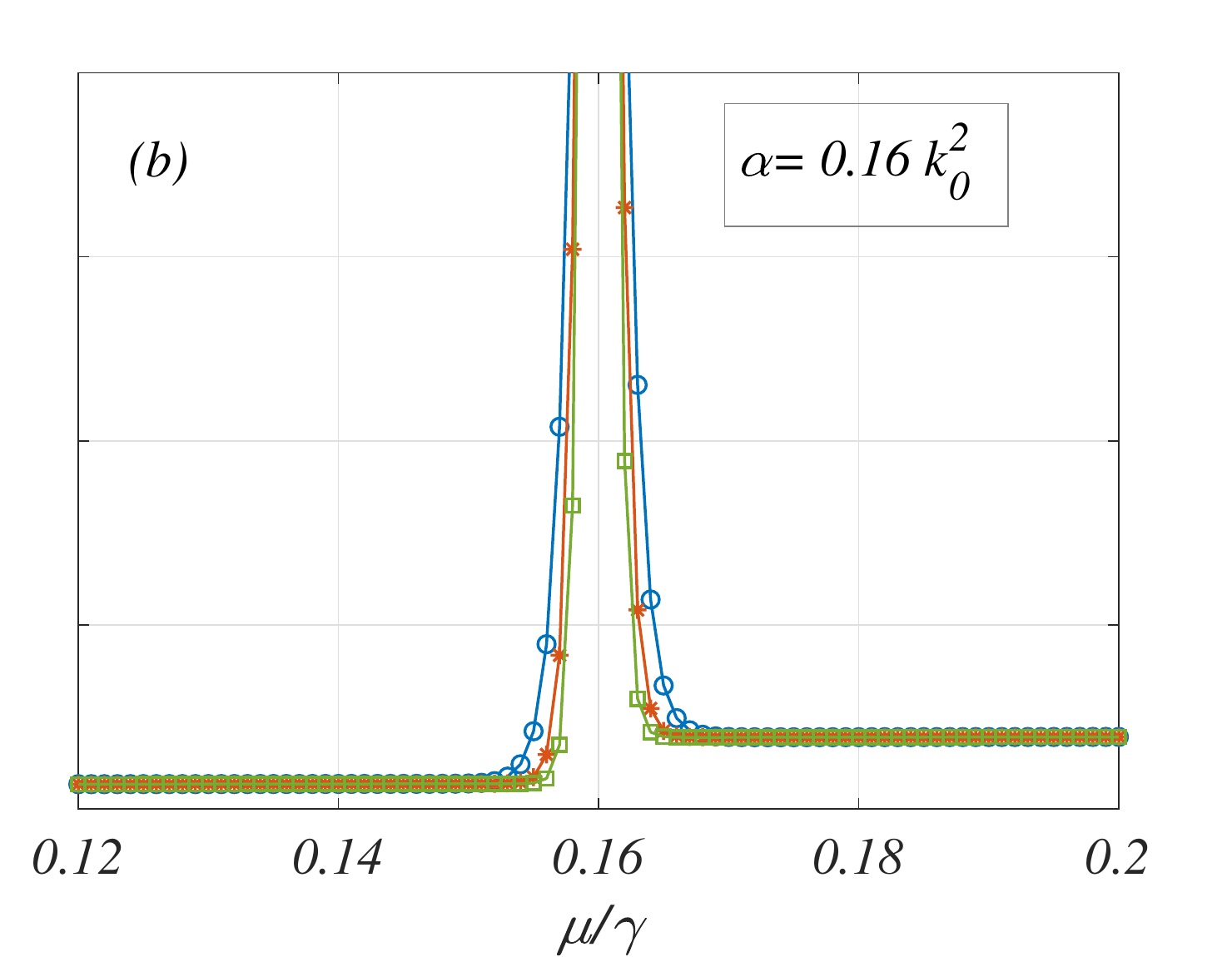}}
\end{minipage}
 \caption{SHC with variation of chemical potential for (a) $\alpha= 0.04k_0^2$ (b) $\alpha= 0.16k_0^2$. The SHC (y-axis) is normalized by $\sigma_0=e(3\hbar ^2/8m^\ast\beta)$. The SHC exhibits resonance when the chemical potential crosses spin degeneracy points i.e., when $\mu/\gamma=\alpha/k_0^2$. The strength of RSOC is taken as $\beta k_0^3=0.4$.  The $\gamma$ is same as in Fig.~(\ref{band})}
 \label{Hall_resonance}
 \end{figure}

{\em Spin-Hall resonance and discussion.}
We now investigate the SHC for the $2$DHG in III-V semiconductor quantum wells with $k^3$-RSOC as a linear response to the applied in-plane electric field along the $x$-direction. The SHC can be evaluated by using the Kubo formula\cite{PhysRevLett126603} as
\begin{eqnarray}
 \sigma_{xy}^{z}&=&\frac{e\hbar}{L_xL_y}\sum_{k,s\neq s'}[f(E_{ks'})-f(E_{ks})]\nonumber\\&\times&\frac{\operatorname{Im}[\langle ks\mid \hat{J}_x^{z}\mid ks' \rangle\langle ks'\mid \hat{v}_y\mid ks\rangle]}{(E_{ks'}-E_{ks})(E_{ks'}-E_{ks}-i\epsilon)}.
\end{eqnarray} 
In the dc limit, $\epsilon$ is set to be zero. The system dimension is denoted by $L_x\times L_y$. The Fermi distribution function is denoted by $f(E_{ks})$ for spin $s$. The spin-current operator is given by $\hat{j}_{x}^{z}=3\hbar/4\{\hat{\sigma}_z,\hat{v}_x\}$.  Note that in general, one should proceed with the Floquet states $\mid k,s;m\rangle$  in the Kubo formula in order to include the possible effects of nearest Floquet side bands. However, here we are only dealing with the zeroth order correction (as the higher-order corrections vanish), hence the effects of Floquet side bands can be safely ignored and we can proceed with the zeroth state $\mid k,s\rangle$. 
The velocity operators are given by $\hat{v}_i=\hbar^{-1}\partial H_{eff}/\partial k_i$ i.e.,
\begin{equation}      
 \hat{v}_x=\frac{\hbar k_x}{m^{\ast}}+\frac{\beta}{\hbar}[3(k_x^2-k_y^2)\,\sigma_y-6k_xk_y\,\sigma_x]+\frac{\alpha\beta}{\hbar}\,\sigma_x
\end{equation} 
and       
\begin{equation}
 \hat{v}_y=\frac{\hbar k_y}{m^{\ast}}+\frac{\beta}{\hbar}[3(k_y^2-k_x^2)\,\sigma_x-6k_xk_y\sigma_y]+\frac{\alpha\beta}{\hbar}\,\sigma_y.
\end{equation}
With the computed matrix elements, we can rewrite the Kubo formula as
\begin{align}\label{shc}
    \sigma^z_{xy}&=& \sigma_0\int \frac{d k}{(2 \pi)^2}\int \cos\theta d\theta [f(E_{k,+})-f(E_{k,-})]\\ \nonumber
    &\times&\frac{ \cos\theta(\alpha^2+3k^4)-\alpha k^2(\sin 3\theta+3 \sin \theta)}{ \mid(k^2-\alpha \sin 2\theta)^2+\alpha^2 \cos^2 2\theta\mid^{3/2}}.
\end{align}
Here, $\sigma_0= (3e/8) (\hbar^2/m^{\ast}\beta)$. The SHC is evaluated numerically by using Eq.(\ref{shc}) and is plotted in Fig.~\ref{Hall_resonance}. 
We normalize $k$ by $k_0$, a typical Fermi momentum corresponding to the hole density. The SHR occurs when the chemical potential becomes equal to $\alpha/k_0^2$, which is exactly the location of the spin degeneracy points in the band dispersion [see Fig.~\ref{band}]. This can also be seen from the denominator of Eq.~(\ref{shc}),which vanishes at the points,$(k_r,\theta_{r1})=(\sqrt{\alpha},\pi/4)$ and $(k_r,\theta_{r2})=(\sqrt{\alpha},5 \pi/4)$, giving the resonance in the SHC. In each panels, three different but very low temperatures are considered as $\beta_{_T}\gamma= 10^3, 1.5\times 10^{3},2\times 10^{3}$, which only affect the broadening of the resonance peaks. Here, $\beta_T=1/k_{B}T$, with $T$ being the absolute temperature and $k_B$, the Boltzmann constant. We also consider two different amplitudes of the light (i.e., location of the degeneracy points), which do not have any significant impact on the resonance phenomena, except for some minor effects on the temperature-dependent width of the resonance peak.   We note that there is a jump in the SHC around the resonance which might be attributed to the asymmetric  contribution of the velocity and spin-current matrix product to the total SHC\cite{comment}. We also check that away from the resonance, the SHC is not constant rather it weakly varies with the chemical potential.   
We use the usual definition of the spin-current operator, even though a modified definition had been used for studying SHR\cite{PhysRevB.75.233306} in electron system and it was noted that the new definition does not affect the SHR phenomena, except for an overall amplitude modification of the SHC.  

Now we comment on the realistic parameter regime for such resonance to occur. At the  degeneracy point, the Fermi momentum $k_F=\sqrt{2\pi n_d}$, can be computed for a typical hole density\cite{PhysRevLett.121.087701} $(n_d\sim 2\times 10^{15}\mathrm{m}^{-2})$ to be around $\sim 10^{8}\mathrm{m}^{-1}$. On the other hand, the typical amplitude of the light field that is used in  experiment is around $eA_0/\hbar\sim (1-10)\times 10^8 \mathrm{m}^{-1}$ corresponding to a photon energy of $\hbar\Omega=0.25$ eV. This suggests that the degeneracy points, $k_F=\sqrt{3}(eA_0/\hbar)$, can be realized in a regular experimental setup\cite{KAESTNER200647}.  

{\em Effects of $k^3$-Dresselhaus type spin-orbit interaction.}

Now we briefly discuss the possible effects of irradiation in presence of $k^3$-DSOC on a heavy-hole gas with RSOC. The DSOC arises due to the lack of bulk inversion symmetry in the system.  This interaction was introduced by Loss in Ref.~[\onlinecite{PhysRevLett.95.076805}] in a study of spin relaxation and decoherence in quantum dots in the presence of a magnetic field. The spin transport was also studied including both the RSOC and DSOC terms\cite{PhysRevB.81.085304}. The total Hamiltonian in presence of RSOC and DSOC can be written as $H=\hbar^2k^2/2m^{\ast}+H_R+H_D$, where the DSOC term is given by $  
   H_D=-(\lambda/2)(k_{-}k_{+}k_{-}\sigma_{+}+k_{+}k_{-}k_{+}\sigma_{-}).
$
 Here, $\lambda$ is the strength of the DSOC. Following the same approach as for RSOC, the linearly-polarized-light-induced correction terms to the Hamiltonian $H$ are given by 
 $-\lambda \alpha(k_x\sigma_x+k_y\sigma_y)/6-\lambda \alpha(k_y\sigma_x+k_x\sigma_y)/3$. The energy eigenvalue in the presence of both the RSOC and the DSOC after incorporating the light induced correction reads as $E_{k,s}=\hbar^2 k^2/2m^{\ast}+s\sqrt{h_1^2+h_2^2}$ with
 \begin{eqnarray}
  h_1&=&-\beta k^3\sin(3\theta)+\eta k\cos\theta-\frac{\alpha\lambda}{3}k\sin\theta,\\
   h_2&=&\beta k^3\cos(3\theta)+\eta k\sin\theta-\frac{\alpha\lambda}{3}k\cos\theta,
 \end{eqnarray}
 where $\eta=\alpha(\beta-\lambda/6)-\lambda k^2$.
It can be seen that the spin splitting still vanishes at $k_r=\sqrt{\alpha (\beta-\lambda/2)/(\beta+\lambda)}$ for $\theta_r=(\pi/4,5\pi/4)$. Therefore, the resonance phenomena observed in SHC should remain unaltered even in the presence of a weak DSOC in the hole system. The only noticeable point here is that now $k_r$ becomes sensitive to the strength of both RSOC and DSOC.\\ 
{\em Conclusion.}
We have shown that linearly polarized light can induce anisotropy in the band structure of a two-dimensional heavy-hole system with $k^3$-RSOC. Most remarkably, the spin-dependent band structure exhibits a pair of additional degeneracy points, which give rise to a resonance in the SHC. The appearance of the resonance in the spin-Hall transport may be an useful approach to achieve a giant SHC. We also discuss the fate of the resonance in the SHC in the presence of a structural-inversion-asymmetry-induced DSOC term and we confirm that the presence of this term only affects the location of the degeneracy points, while keeping the resonance phenomena unaffected. Finally, we would also like to comment that like the case of a two-dimensional electron system with $k$-RSOC, the application of circularly polarized light does open a gap between spin branches in our case.  Therefore, the application of circularly polarized light will not be useful to achieve the spin-Hall resonance.\\ 
{\em Acknowledgements}
The authors thank Carsten Timm, Shyamal Biswas, Tarun K. Ghosh and A. A. Zyuzin for useful discussion. A.B. acknowledges financial support by the Deutsche
Forschungsgemeinschaft. SFI acknowledges financial support by the Academy of Finland Grant No. 308339.
\bibliography{shr_ref}

\begin{thebibliography}{73}
\bibitem [{com()}]{comment}%
  \BibitemOpen
  \href@noop {} {\enquote {\bibinfo {title} {The jump can be understood by
  analyzing the integrand at $\theta_r=\pi/4$ as
  $[f(e_{k,+})-f(e_{k,-})](3k^2-\alpha)sgn(k^2-\alpha)/\mid k^2-\alpha\mid^2$
  which is not symmetric across the resonance.}}\ }\BibitemShut {NoStop}%
\end{thebibliography}
\end{document}